\documentstyle[twoside,multicol,epsfig]{article}
\catcode`\@=11
\long\def\@makefntext#1{
\protect\noindent \hbox to 3.2pt {\hskip-.9pt  
$^{{\footnotesize\@thefnmark}}$\hfil}#1\hfill}		

\def\@makefnmark{\hbox to 0pt{$^{\@thefnmark}$\hss}}	
	
\def\ps@myheadings{%
    \let\@oddfoot\@empty\let\@evenfoot\@empty
    \def\@evenhead{\footnotesize\it\leftmark\hfil}
    \def\@oddhead{\hfil{\footnotesize\it\rightmark}}
    \let\@mkboth\@gobbletwo
    \let\sectionmark\@gobble
    \let\subsectionmark\@gobble
    }

\oddsidemargin=\evensidemargin
\newcounter{sectionc}\newcounter{subsectionc}\newcounter{subsubsectionc}
\renewcommand{\section}[1] {\vspace{14pt}\addtocounter{sectionc}{1}
\setcounter{subsectionc}{0}\setcounter{subsubsectionc}{0}\noindent 
	{\bf\thesectionc. #1}\par\vspace{8pt}}
\renewcommand{\subsection}[1] {\vspace{14pt}\addtocounter{subsectionc}{1}
   \setcounter{subsubsectionc}{0}\noindent 
   {\bf\thesectionc.\thesubsectionc. {\kern1pt \bfit #1}}\par\vspace{8pt}}
\renewcommand{\subsubsection}[1] {\vspace{14pt}
    \addtocounter{subsubsectionc}{1}
	\noindent{\thesectionc.\thesubsectionc.\thesubsubsectionc.
	{\kern1pt \it #1}}\par\vspace{8pt}}
\newcommand{\nonumsection}[1] {\vspace{14pt}\noindent{\bf #1}
	\par\vspace{8pt}}

\newcounter{appendixc}
\newcounter{subappendixc}[appendixc]
\newcounter{subsubappendixc}[subappendixc]
\renewcommand{\thesubappendixc}{\Alph{appendixc}.\arabic{subappendixc}}
\renewcommand{\thesubsubappendixc}

\renewcommand{\appendix}[1] {\vspace{14pt}
        \refstepcounter{appendixc}
        \setcounter{figure}{0}
        \setcounter{table}{0}
        \setcounter{lemma}{0}
        \setcounter{theorem}{0}
        \setcounter{corollary}{0}
        \setcounter{definition}{0}
        \setcounter{equation}{0}
        \renewcommand{\thefigure}{\Alph{appendixc}.\arabic{figure}}
        \renewcommand{\thetable}{\Alph{appendixc}.\arabic{table}}
        \renewcommand{\theappendixc}{\Alph{appendixc}}
        \renewcommand{\thelemma}{\Alph{appendixc}.\arabic{lemma}}
        \renewcommand{\thetheorem}{\Alph{appendixc}.\arabic{theorem}}
        \renewcommand{\thedefinition}{\Alph{appendixc}.\arabic{definition}}
        \renewcommand{\thecorollary}{\Alph{appendixc}.\arabic{corollary}}
        \renewcommand{\theequation}{\Alph{appendixc}.\arabic{equation}}
        \noindent{\bf Appendix \theappendixc #1}\par\vspace{5pt}}
\newcommand{\subappendix}[1] {\vspace{14pt}
        \refstepcounter{subappendixc}
        \noindent{\bf Appendix \thesubappendixc. {\kern1pt \bfit #1}}
	\par\vspace{8pt}}
\newcommand{\subsubappendix}[1] {\vspace{14pt}
        \refstepcounter{subsubappendixc}
        \noindent{\rm Appendix \thesubsubappendixc. {\kern1pt \it #1}}
	\par\vspace{8pt}}

\newcommand{\textlineskip}{\baselineskip=13pt}
\newcommand{\smalllineskip}{\baselineskip=10pt}

\newcommand{\copyrightheading}[1]
	{\vspace*{-2.5cm}\smalllineskip{\flushleft
	{\footnotesize {\it Preprint submitted to VUV-13}}\\
	 }}

\renewenvironment{thebibliography}[1]		
	{\small\baselineskip=11pt
	 \frenchspacing
	 \begin{list}{\arabic{enumi}.}
        {\usecounter{enumi}\setlength{\parsep}{0pt}    
	 \setlength{\leftmargin 12.7pt}{\rightmargin 0pt}
         \setlength{\itemsep}{0pt} \settowidth
	{\labelwidth}{#1.}\sloppy}}{\end{list}}

\newcounter{itemlistc}
\newcounter{romanlistc}
\newcounter{alphlistc}
\newcounter{arabiclistc}

\newcommand{\fcaption}[1]{
        \refstepcounter{figure}
        \setbox\@tempboxa = \hbox{\small Fig.~\thefigure. #1}
        \ifdim \wd\@tempboxa > 5in
           {\begin{center}
        \parbox{2.8in}{\small\baselineskip=11pt Fig.~\thefigure. #1}
            \end{center}}
        \else
             {\begin{center}
             {\small Fig.~\thefigure. #1}
              \end{center}}
        \fi}

\newcommand{\tcaption}[1]{
        \refstepcounter{table}
        \setbox\@tempboxa = \hbox{\small Table~\thetable. #1}
        \ifdim \wd\@tempboxa > 5in
           {\begin{center}
        \parbox{5in}{\small\baselineskip=11pt Table~\thetable. #1}
            \end{center}}
        \else
             {\begin{center}
             {\small Table~\thetable. #1}
              \end{center}}
        \fi}

\def\@citex[#1]#2{\if@filesw\immediate\write\@auxout
	{\string\citation{#2}}\fi
\def\@citea{}\@cite{\@for\@citeb:=#2\do
	{\@citea\def\@citea{,}\@ifundefined
	{b@\@citeb}{{\bf ?}\@warning
	{Citation `\@citeb' on page \thepage \space undefined}}
	{\csname b@\@citeb\endcsname}}}{#1}}

\newif\if@cghi
\def\cite{\@cghitrue\@ifnextchar [{\@tempswatrue
	\@citex}{\@tempswafalse\@citex[]}}
\def\citelow{\@cghifalse\@ifnextchar [{\@tempswatrue
	\@citex}{\@tempswafalse\@citex[]}}
\def\@cite#1#2{{$\null^{#1}$\if@tempswa\typeout
	{IJCGA warning: optional citation argument 
	ignored: `#2'} \fi}}

\def\pmb#1{\setbox0=\hbox{#1}
	\kern-.025em\copy0\kern-\wd0
	\kern.05em\copy0\kern-\wd0
	\kern-.025em\raise.0433em\box0}

\def\fnt#1#2{\footnotetext{\kern-.3em
	{$^{\mbox{\scriptsize #1}}$}{#2}}}

\font\tenit=cmti10 

\font\bfit=cmbxti10 at 10pt

\font\eightit=cmti8

\def\itlatex{\tenit L\kern-.30em\raise.4ex\hbox{\eightit A}\kern-.14em 
T\kern-.1667em\lower.7ex\hbox{E}\kern-.125em X} 

\def\bsc{{\sc a\kern-7pt\sc a}}
\def\bflatex{\bf L\kern-.30em\raise.3ex\hbox{\bsc}\kern-.18em
T\kern-.1667em\lower.7ex\hbox{E}\kern-.125em X} 

\begin{document}

\setlength{\textheight}{8.8truein}     

\thispagestyle{empty}

\markboth{J. D. Denlinger {\it et al.}}
{J. D. Denlinger {\it et al.}}

\textlineskip
\setcounter{page}{1}

\copyrightheading{Vol.~0, No.~0 (2000) 00--00} 

\vspace*{0.55truein}


\centerline{\large\bf BULK BAND GAPS IN DIVALENT HEXABORIDES:}
\centerline{\large\bf A SOFT X-RAY EMISSION STUDY}

\vspace*{0.04truein}

\centerline{J. D. DENLINGER}
\vspace*{0.0215truein}
\centerline{\it Advanced Light Source, Lawrence Berkeley National Lab}
\baselineskip=11pt
\centerline{\it Berkeley, CA 94720, USA}
\centerline{\it E-mail: JDDenlinger@lbl.gov}
\vspace*{0.2truein}

\centerline{G.-H. GWEON, J. W. ALLEN}
\vspace*{0.0215truein}
\centerline{\it Randall Laboratory of Physics, University of
Michigan}
\centerline{\it Ann Arbor, MI 48109-1120, USA}
\vspace*{0.2truein}

\centerline{A. D. BIANCHI\cite{lanl}, Z. FISK}
\vspace*{0.0215truein}
\centerline{\it National High Magnetic Field Lab, Florida State
University}
\centerline{\it Tallahassee, FL 32306, USA}

\vspace*{0.25truein}

\vspace*{0.25truein}
\abstract{Boron K-edge soft x-ray emission and absorption
are used to address the fundamental question of whether divalent
hexaborides are intrinsic semimetals or defect-doped bandgap insulators.  
These bulk sensitive measurements, complementary and consistent with
surface-sensitive angle-resolved photoemission experiments, confirm the
existence of a bulk band gap and the location of the chemical
potential at the bottom of the conduction band. }{}

\vspace*{0.35truein}\textlineskip

\def\EF{$E_{\rm F}$}
\def\kF{$k_{\rm F}$}
\def\hv{h$\nu$}
\def\kx{k$_x$}
\def\ky{k$_y$}
\def\kz{k$_z$}
\def\G{$\Gamma$}
\def\sbar{$\overline{\rm S}$}
\def\gbar{$\overline{\Gamma}$}
\def\gradv{$\vec\nabla V$}
\def\kpar{$k_{\parallel}$}
\def\deg{$^{\circ}$}
\def\A-1{$\AA^{-1}$}
\def\B6{B$_6$}
\def\CaLaB6{Ca$_{0.995}$La$_{0.005}$B$_6$}
\def\(1-d){$_{1-\delta}$}
\def\1+d{$_{1+\delta}$}
\def\Å{$\approx$}
\def\site{$cite$}

\begin{multicols}{2}
\section{Introduction}
\noindent

The discovery of weak itinerant ferromagnetism in certain divalent
hexaborides\cite{Young99,Vonlanthen00,Ott00,Terashima00} provides strong
motivation to determine the underlying electronic structure giving rise
to the metallic carriers.  One possibility predicted by LDA band
calculations\cite{Hasegawa79,Massidda97}, and supported by the
interpretation given to magneto-oscillatory
studies\cite{Goodrich98,Aronson99}, is a semi-metallic band overlap at
the X-point of the cubic Brillouin zone, the absence of which would
render stoichiometric material to be insulating.  Several theoretical
discussions\cite{Zhitomirsky99,Balents00,Barzykin00} presume the
existence of such an overlap.  As summarized
elsewhere\cite{Denlinger-prl01} we have recently given conclusive
experimental proof that there is instead an X-point gap.  This result
eliminates all models that assume overlap and so may favor a
ferromagnetic dilute electron gas picture\cite{Ceperley99}, but it also
forces the consideration of boron vacanies as the origin of the electrons
observed in nominally stoichiometric divalent materials, possibly
favoring a picture in which the magnetic moments are carried by boron
vacancies\cite{Monnier-condmat01}.

The X-point gap was first observed in angle resolved photoemission
(ARPES) of Eu\B6\ and Sr\B6\ and ascribed by us initially only to the
surface region probed in ARPES\cite{Denlinger-condmat00}.  Strong
motivation to reinterpret the ARPES result as showing a bulk
gap\cite{aps01} was provided by a recent band
calculation\cite{Tromp-condmat00} that includes a GW self energy
correction and predicts Ca\B6\ to have the X-point band gap of 0.8 eV
similar to that measured by ARPES.  Results from bulk sensitive soft
x-ray emission and absorption spectroscopy (SXE and XAS, respectively)
showing the gap are an essential part of our experimental proof.  This
paper gives important aspects of the SXE/XAS results and analysis that
were not presented previously.

\vspace*{-4pt}
\section{Experimental}
\noindent

Single crystal hexaboride samples were grown from an aluminum flux
using powders prepared by boro-thermally reducing cation oxides, a method
shown to yield high quality with regard to both  structure and chemical
composition\cite{Ott97}.   Soft x-ray emission and absorption
experiments were performed at the ALS Beamline 8.0.1 using the
Tennessee/Tulane grating spectrometer.  The experimental emission and
absorption spectral resolutions were \Å0.35 eV and \Å0.1 eV,
respectively.  SXE, measured with a 1500 line/mm grating for fixed photon
energy excitation at and above the B K-edge threshold, is used as a probe
of the occupied boron partial density of states for dipole-allowed
transitions back to the B $1s$ core-level, i.e. $p$-states.  X-ray
absorption, a probe of the unoccupied states, was measured both with
total electron yield (TEY) as a function of photon energy and also with
partial fluorescence yield (PFY) with the detection window covering the
entire valence band emission.  Differences between TEY and PFY signals
arise from differing attenuation lengths and the experimental geometry
which was set to 60\deg\ incidence excitation and 30\deg\ grazing
emission relative to the sample surface.
Absolute PFY energies were calibrated to published TEY
spectra of the hexaborides\cite{Jia96} and SXE spectra were calibrated to
the excitation energy via the presence of elastic scattering in the
emission spectra.  Previous very early work on the hexaborides using
electron-excited soft x-ray emission and thin film absorption
measurements\cite{Lyakhovskaya70,Okusawa82} do not provide clear enough
spectral detail at threshold to address the band gap issues that have
only recently been raised. 

\section{Results}
\noindent

Fig. 1 shows a representative data set of soft x-ray emission and
absorption at the boron K-edge for the divalent hexaborides.  Very
similar data to this example from a cleaved Yb\B6\ sample was also
obtained for Ca\B6, Sr\B6, and Eu\B6.  Fig. 1(a) compares the two methods
of measuring x-ray absorption, TEY and PFY.  TEY, a measurement of
the sample current, exhibits a sharp peak at $\approx$194 eV
corresponding to a well-known B $1s$$\rightarrow$$2p$ $\pi$*
transition\cite{Jia96}, which in part arises from surface layer oxidation
of the air-cleaved sample. Also the absolute TEY signal exhibits a high
background (removed in Fig. 1(a)) with declining slope due to the
presence of lower energy absorption edges. 

In contrast, the PFY signal, a measure of valence emission
intensities integrated over the energy window shown in Fig. 1(b), does not
exhibit the sharp TEY absorption peak due to a greater bulk
emission sensitivity than the excitation depth which rapidly
decreases while scanning through the absorption resonance. Also the
valence band PFY signal inherently has zero pre-threshold intensity and
hence is preferred over TEY for careful measurement of the threshold
region.  The PFY spectrum for Yb\B6\ shows a weak step-like threshold
onset at 187.1 eV which is highlighted by a logarithmic scale in the
inset to Fig. 1(a).  TEY also shows an intensity cusp at this energy, but
only for cleaved surfaces with minimal surface contamination.  We
interpret this threshold onset in the PFY spectrum as the energy position
of the chemical potential which we define at the half-step intensity.
This location of the chemical potential or Fermi-edge (\EF) in the
conduction band is immediately important for electronic structure models
and is consistent with negative Hall coefficient measurements indicating
the presence of electron (and not hole) carriers in the
bulk\cite{Fisk79,Tarascon80}.

Valence band emission spectra, shown in Fig. 1(b) were acquired at selected
photon energies indicated by arrows in Fig. 1.  An elastic peak present in
the emission spectra is resonantly
enhanced at the B $1s$$\rightarrow$$2p$ absorption peak and is used for
calibration of the SXE energy scale to that of TEY and PFY spectra.  
Excitation at the selected peaks in the PFY/TEY spectra and far
above threshold show similar valence band emission profiles with small
variations in the relative intensities of at least six discernable peaks
and shoulders.  Threshold excitation on the other hand produces much
larger variation in the relative intensities (and energies) of the
different valence emission peaks (discussed in the next section).  The
elastic peak is also observed to be enhanced at threshold, thus providing
a distinct marker of \EF\ for all the emissionspectra.  The non-threshold
SXE spectra, on the other hand, exhibit a strong non-metallic decay of
intensity approaching \EF, in contrast to the weak step-onset in the PFY
spectrum which implies a small occupancy of states at \EF.  A lack of
clear SXE detection of such density of states near \EF\ could
be due to too little occupancy, very weak boron $p$-character, and/or a
poorer SXE resolution than the occupied bandwidth.


\section{Comparison to band theory}
\noindent

A detailed comparison of the Yb\B6\ SXE and PFY spectra to both LDA and GW 
calculations, presented in Fig. 2, reveals a clear distinction between the
band-overlap and band gap electronic structure models.  Available
partial density of states (DOS) calculations exist in the literature for
Sr\B6\ only\cite{Massidda97} and the GW calculation was performed for
Ca\B6\cite{Tromp-condmat00}.  However, since experimental SXE bandwidths
of all hexaborides are very similar and the theoretical bandwidths for a
given calculation method are also similar between different hexaborides,
the comparison of experiment to theory involving different divalent cations
is justified.

Fig. 2(a) shows the combined SXE (194.1 eV excitation) and PFY spectra
with the chemical potential shifted to 0 eV and with comparison to the
theoretical LDA $p$-DOS for Sr\B6.  The relative SXE and PFY intensities
have been scaled to match the theoretical DOS amplitudes above and below
\EF. Boron $s$-DOS and cation $d$-DOS, not probed by these measurements,
are strongest at -8 eV and above +5 eV, respectively. The corresponding
$k$-resolved LDA band structure (for Ca\B6) along
\G-X for these DOS, exhibiting a small band overlap at the X-point between
the valence and conduction bands, is plotted in Fig. 2(b).  A clear
discrepancy between the energy position of the LDA occupied and unoccupied
DOS with the experimental spectra is seen. 

A rigid shift of the theoretical DOS to lower energy provides a better
agreement with experiment. The overall occupied boron-block bandwidth and
relative peak amplitudes are in good agreement with the LDA $p$-DOS.
However, independent energy shifts of the occupied and unoccupied states
(-0.8 eV and -0.5 eV, respectively) is required to make best alignment to
the 2.5-5 eV SXE and 3 eV PFY peaks (Fig. 2(c)).  This suggests a
relatively larger experimental energy separation between the valence and
conduction bands than predicted by LDA and thus no X-point band overlap.  

Fig. 2(d) shows the GW bands along \G-X for Ca\B6\ exhibiting a band gap
at the X-point.  The GW band calculation predicted a 10\% expansion of
the boron-block bandwidth compared to LDA.  However this larger GW
bandwidth is not consistent with the bandwidths measured here by SXE or by
ARPES\cite{Denlinger-prl01} which are in better agreement with LDA. 
Hence for Fig. 2(d) the GW energy scale has been multiplied by 0.9 to
match the LDA bandwidth, and then shifted to lower energy placing
\EF\ at the bottom of the X-point conduction band consistent with the PFY
measurement. 

Since the divalent hexaboride conduction band minimum resides at the
X-point, threshold excitation into these unoccupied states is
expected to exhibit X-point $k$-selectivity in the emission
process due to the lack of intermediate scattering paths to other lower
energy $k$-points. Indeed, a favorable one-to-one correspondence can be
made between the X-point band energies of the modified GW band
calculation in Fig. 2(d) to the dominant peaks labeled ($a-e$) in the
threshold-excited SXE spectrum plotted in Fig. 2(e). Most important for
our goal of distinguishing band-overlap versus band-gap electronic
structure is the identification of peak $a$ as the boron-block valence band
maximum at $\approx$ 1 eV below \EF.  The threshold-excited Yb\B6 SXE
spectrum thus provides a direct quantification of the bulk bandgap to be
1.0 eV minus the energy that the conduction band dips below \EF, thus
providing a firm basis for the reasoning set forth
in\cite{Denlinger-prl01}.

\section{Discussion}
\noindent

These soft x-ray measurements verifying the existence of a bulk band gap
are complementary to our surface-sensitive ARPES measurements of the \G-X
band structure\cite{Denlinger-prl01,Denlinger-condmat00} which
provide a more detailed view of the X-point gap.  The position of the
chemical potential at the bottom of the conduction band is also
qualitatively consistent with ARPES measurements of an electron pocket at
the X-point.  While excess electron concentrations at the surface can be
explained by band bending and charge redistribution effects, excess
electron carriers in the bulk forces one to confront the issue of
off-stoichiometry defects, i.e. electron counting and band filling of the
hexaboride band structure with the presence of a band gap predicts
divalent materials to be insulators.  A rigid covalently bonded boron
sublattice with mobile cations is suggestive of cation vacancies only,
i.e. hole carriers, inconsistent with Hall coefficient  measurements.  The
presence of boron vacancies on the other hand appears to be a more
natural source of excess electrons and such defects have even been
recently shown theoretically to possess magnetic moments which may be
highly relevant to the origin of the anomalous ferromagnetism in La-doped
Ca\B6\ and other divalent hexaboride systems\cite{Monnier-condmat01}.

We have also found from a combination of SXE/PFY and ARPES
spectra\cite{Denlinger-prl01,Mo-sces01} that the X-point gap is absent
for trivalent La\B6\ and mixed valent Sm\B6\, implying a non-trivial
transformation from divalency to trivalency.  Additionally a distinct
anomalous deviation from Vegard's law exists in the Ca$_{1-x}$La$_x$\B6\
series.  Rather than a monotonic increase in lattice constant expected
from the larger La$^{3+}$ size, the lattice parameter at first shrinks to
a minimum at $\approx$10\% La doping before increasing again at higher
La-doping\cite{Bianchi-tbp}.  This critical La concentration may be
correlated with a crossover from band-gap to band-overlap X-point
electronic structure.

\section{Summary}
\noindent

Boron K-edge soft x-ray emission and absorption have been used to probe
boron occupied and unoccupied partial density of states of the divalent
hexaborides, with data from Yb\B6\ used as an example.  The chemical
potential is identified to be located at the bottom of the conduction
states in the absorption spectra and comparison of the emission spectra to
calculated LDA density of states is used to identify the energy of
the valence band maximum to be $\approx$1 eV below the chemical potential.
This result establishes the existence of a bulk X-point band gap consistent
with recent GW band calculations and rules out the band overlap model as a
starting point for theories to explain the novel ferromagnetism in doped
and undoped hexaborides.


\nonumsection{Acknowledgments}
\noindent
This work was supported at U. of Michigan by the U.S. DoE under Contract No. 
DE-FG02-90ER45416 and by the U.S. NSF Grant No. DMR-99-71611.  The ALS is
supported by the U.S. DoE under contract No. DE-AC03-76SF00098.

\nonumsection{References}

\hfill\break




\vspace*{13pt}
\centerline{\psfig{figure=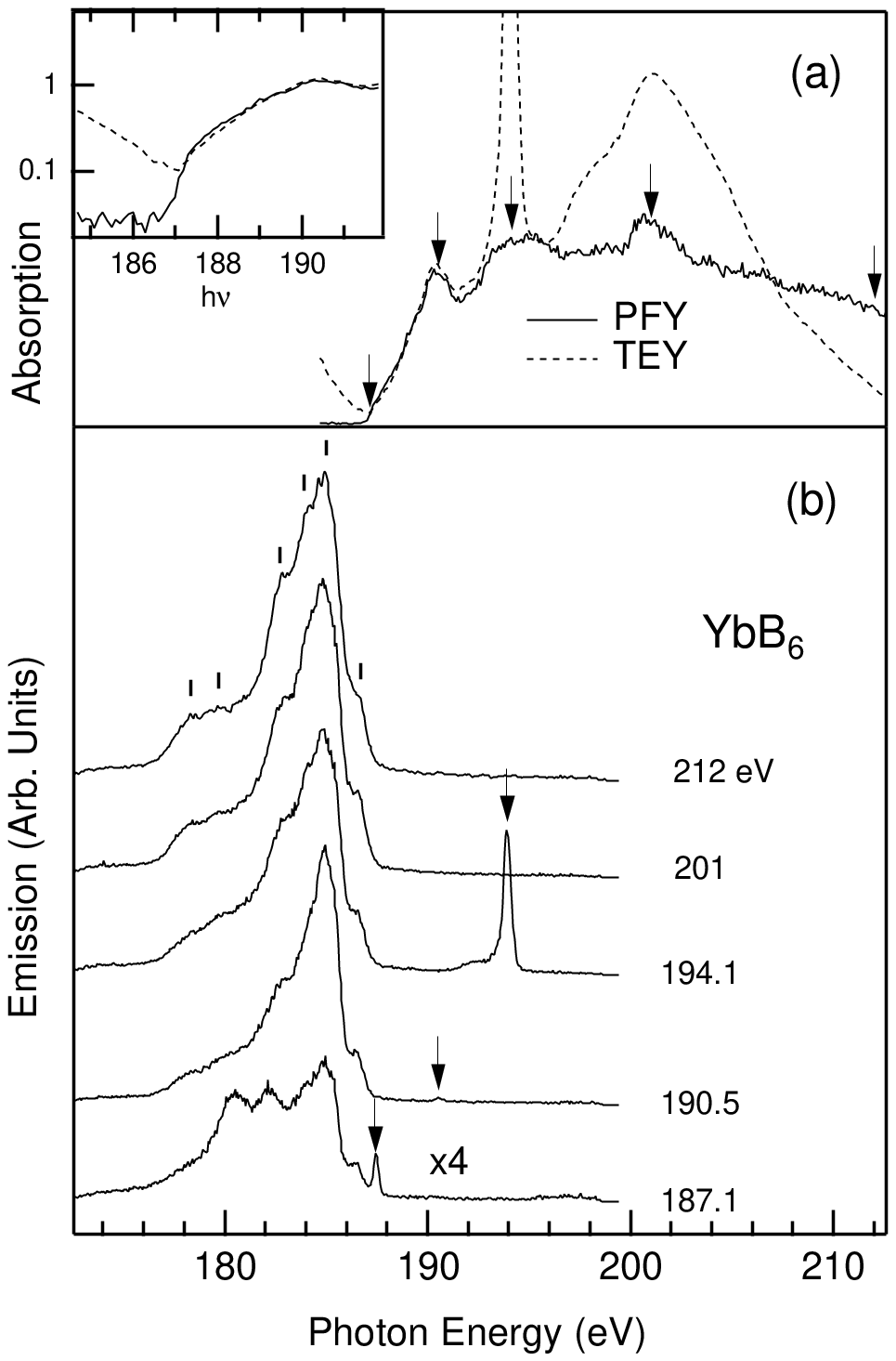, width=3.5in}}
\fcaption{Soft x-ray absorption (TEY, PFY) and emission (SXE) boron K-edge
data set for Yb\B6.  Arrows and values indicate the excitation energies.
The logarithmic intensity scale of the inset highlights the step intensity
onset of the PFY signal.}
\vspace*{13pt}

\centerline{\psfig{figure=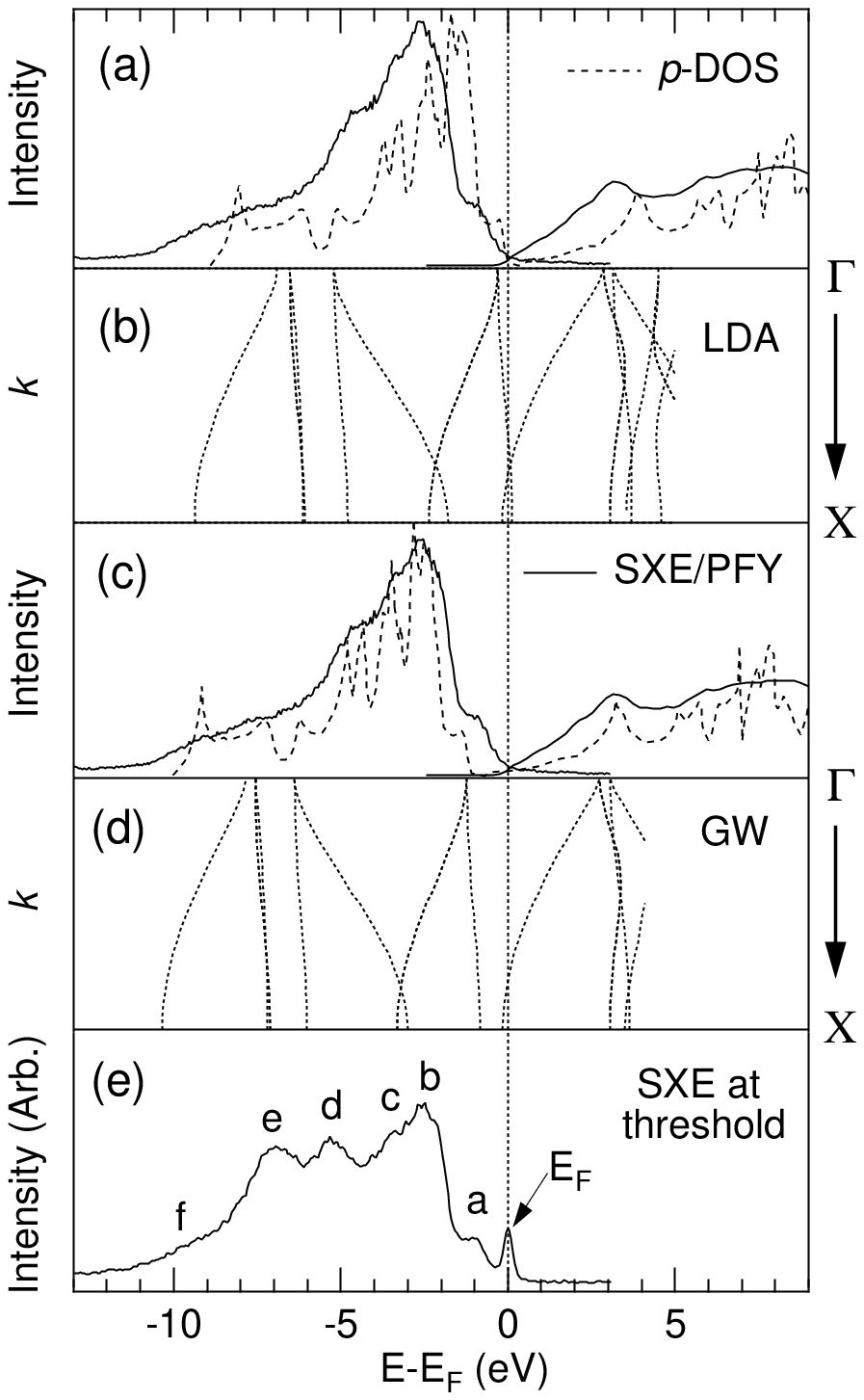, width=3.25in}}
\fcaption{(a) Comparison of Yb\B6\ SXE and PFY to LDA boron $p$-DOS for
Sr\B6, (b) LDA band structure along \G-X exhibiting band overlap at X,
(c) comparison of  SXE and PFY to energy shifted boron $p$-DOS, (d) GW
band structure along \G-X exhibiting a band gap, (e) Yb\B6\ SXE excited
at threshold exhibiting X-point $k$-selectivity of peaks ($a$-$e$).}
\vspace*{13pt}

\end{multicols}
\end{document}